\documentclass[aps,pra,showpacs,preprintnumbers,amsmath,amssymb,footinbib]{revtex4}
\usepackage{graphicx,epsfig}
\usepackage{subfig}
\usepackage{bm}
\usepackage{xcolor}
\usepackage[breaklinks=true,colorlinks,citecolor=blue,linkcolor=blue,urlcolor=blue]{hyperref}

\chardef\us=`\_

\def\noi{\noindent}
\def\bc{\begin{center}}
\def\ec{\end{center}}
\newcommand{\avec}{ \vec A}

\newcommand{\bb}{\vec B}

\newcommand{\en}{{\boldsymbol{\hat{n}}}}

\newcommand{\be}{\begin{equation}}
\newcommand{\ee}{\end{equation}}
\newcommand{\ber}{\begin{eqnarray}}
\newcommand{\eer}{\end{eqnarray}\noi}
\begin{document}
\title{Self-consistent Nonlinear Force-free Field Reconstruction from Weighted Boundary Conditions}

\author{Alpha Mastrano$^{1}$}\email{alpha.mastrano@sydney.edu.au}
\author{Kai E. Yang$^{1}$}
\author{Michael S. Wheatland$^{1}$}
\affiliation{$^{1}$Sydney Institute for Astronomy, School of Physics, University of Sydney, NSW 2006, Australia}

\date{\today}

\begin{abstract}
Vector magnetogram data are often used as photospheric boundary conditions for force-free coronal magnetic field extrapolations. In general, however, vector magnetogram data are not consistent with the force-free assumption. In this article, we demonstrate a way to deal with inconsistent boundary data, by generalizing the ``self-consistency procedure" of \cite{wheatlandregnier09}. In that procedure, the inconsistency is resolved by an iterative process of constructing two solutions based on the values of the force-free parameter $\alpha$ on the two polarities of the field in the boundary (the P and N polarities), and taking uncertainty-weighted averages of the boundary $\alpha$ values in the P and N solutions. When the $\alpha$ values in the P and N regions are very different, the self-consistent solution may lose high $\alpha$ values from the boundary conditions. We show how, by altering the weighting of the uncertainties in the P or N boundary conditions, we can preserve high $\alpha$ values in the self-consistent solution. The weighted self-consistent extrapolation method is demonstrated on an analytic bipole field and applied to vector magnetogram data taken by the \emph{Helioseismic and Magnetic Imager} (HMI) instrument for NOAA active region AR 12017 on 2014 March 29.

\end{abstract}

\maketitle
 

\section{Introduction}
     \label{Introduction} 

The magnetic field in the solar corona cannot be measured directly. The vector magnetic field at the photosphere, however, can be inferred from spectropolarimetric measurements of lines in the low atmosphere. The photospheric measurements can then, in principle, be used as boundary conditions to extrapolate the coronal magnetic field. A common simplifying assumption is that the coronal field is force-free, \emph{i.e.} the Lorentz force is zero. The problem then reduces to solving a boundary value problem, given by two vector partial differential equations for the magnetic field $\bb$, namely the physical requirement that the magnetic field is divergence-free

\be \nabla \cdot \bb = 0,\ee \label{eq:div}
and the force-free condition

\be \nabla \times \bb = \alpha \bb, \label{eq:ff}\ee 
where $\alpha$ is the scalar force-free parameter (which in general is a function of position). Taking the divergence of Equation \ref{eq:ff} allows us to recast it into

\be \bb \cdot \nabla\alpha=0,\label{eq:ff2}\ee
\emph{i.e.}, $\alpha$ is constant along field lines.

There are several methods available for extrapolating a nonlinear force-free field (NLFFF) in the solar corona from photospheric boundary conditions [for summaries and comparisons, see, \emph{e.g.}, \cite{schrijveretal06,schrijveretal08}, \cite{derosaetal09}, \cite{wiegelmannsakurai12}, \cite{regnier13}, and \cite{wiegelmannpetrieriley17}]. One class of methods uses the Grad-Rubin approach, which solves for $\bb$ and $\alpha$ via an iterative procedure, and uses $\alpha$ and the normal component of $\bb$ at the photosphere as boundary conditions.

Because $\alpha$ is constant along field lines, Grad-Rubin methods only require as boundary condition the value of $\alpha$ over one magnetic polarity, for field lines connected to the lower boundary at both ends. If the values of $\bb$ and $\alpha$ are available over an entire solar active region, the boundary value problem is over-specified. One needs to make a choice as to which set of $\alpha$ values to take as boundary condition --- those from the positive magnetic polarity region or the negative region. The results of the two possible reconstructions are referred to as the P solution and the N solution, respectively. In principle, if the boundary conditions are consistent with a force-free field, this over-specification of the problem is not an issue and the two solutions are equal. However, in practice the P and N solutions can differ significantly, both qualitatively and quantitatively: they may have substantially different energies, and one solution may show field structures and field line connectivities that the other lacks. This occurs because the magnetic field on the photosphere, which is used to derive the boundary conditions for NLFFF reconstructions, is not force-free \citep{gary01}. One approach to deal with the inconsistency of the boundary conditions with the force-free model, developed by \citet{wiegelmanninhestersakurai06}, is ``preprocessing," which alters the magnetogram data to meet necessary conditions for the existence of a force-free solution. 

A different approach is to take the data as given, but allow the boundary values of $\bb$ and $\alpha$ to change during the NLFFF extrapolation process \citep{wiegelmannsakurai12}. This procedure is used in the modified Grad-Rubin code (XTRAPOL) of \citet{amarialy10} and the ``self-consistency" procedure of \citet{wheatlandregnier09}, which uses the Grad-Rubin current-field iteration code CFIT of \cite{wheatland07}. The \cite{wheatlandregnier09} method calculates P and N solutions from the two sets of boundary conditions on $\alpha$, and then averages the values of $\alpha$ for the two solutions at every point on the boundary, taking into account the uncertainties in $\alpha$. The resulting map of $\alpha$ allows two new P and N solutions to be constructed, and then the cycle is iterated until a self-consistent set of values is obtained, \emph{i.e.} the P and N solutions are the same.

The self-consistency procedure uses the uncertainties in the values of $\alpha$ on the boundary (calculated from $\bb$-measurement uncertainties) as weightings when calculating the average value of $\alpha$ at the end of each cycle. When applied with the observational uncertainties, the procedure can lead to a loss of information regarding areas with high values of $\alpha$, resulting in a self-consistent solution with a smooth distribution of boundary values of $\alpha$ compared to the observed boundary conditions. This is because, when the uncertainties in $\alpha$ on the two polarities are approximately equal, each self-consistency cycle takes a simple average of the values of $\alpha$ from the P and N polarities. For example, \cite{kleintetal18} presented reconstructions using the self-consistency procedure of the coronal field associated with active region (AR) 12017 on 2014 March 29. They found the P and N solutions to be quite different: the P solution included a highly twisted flux rope along the magnetic neutral line, and this was absent in the N solution. In this particular case, the positive polarity field has regions with high values of $\alpha$, which do not have counterparts in the negative polarity regions \citep{kleintetal18}. In successive self-consistency cycles, the large values of $\alpha$ at certain boundary points in the P solution are averaged with small values of $\alpha$, leading to a decrease in $\alpha$ at those locations. Therefore, while the procedure constructs a final NLFFF that is consistent with the boundary conditions on both polarities, that field may be close to potential, and have magnetic field lines which lack resemblance to structures observed in extreme ultraviolet (EUV) observations.

In this article, we present a modification to the self-consistency procedure that can generate a self-consistent NLFFF that agrees better with observations by weighting the uncertainties in the values of $\alpha$ in the boundary conditions towards areas which are considered significant. The article is structured as follows. In Section \ref{method} we describe our modification to the self-consistency procedure as implemented by the CFIT method. In Section \ref{analytic}, we demonstrate the method using an analytic bipolar field. In Section \ref{ar12017} we apply our method to the NLFFF extrapolation of active region AR 12017 using the data obtained by the \emph{Helioseismic and Magnetic Imager} (HMI) instrument aboard the \emph{Solar Dynamics Observatory} (SDO) spacecraft. We compare our NLFFF reconstructions to the EUV observations taken by the \emph{Atmospheric Imaging Assembly} (AIA) instrument aboard SDO. In Section \ref{conclusions} we summarize our results.


\section{Data and Methods}
      \label{method}

For the solar coronal field reconstructions in this article, we use the HMI vector magnetogram product called Space-Weather HMI Active Region Patches (SHARPs) \citep{sun13arxiv,hoeksemaetal14}. The NLFFF code used here (CFIT) works in a Cartesian geometry. In application to the SHARP data, which are in a cylindrical equal-area projection, we assume the magnetic vector field values are on a Cartesian grid. We take the photosphere to be in the $x$-$y$ plane and the positive $z$-direction is the radial direction away from the Sun, and we assume $B_x=B_\phi$, $B_y=-B_\theta$, and $B_z=B_r$, where $B_r$, $B_\theta$, and $B_\phi$ are the field components in the SHARP data. The force-free parameter $\alpha_0$ at $z=0$, which we require as boundary condition, is obtained from

\be \alpha_0=\frac{1}{B_z}\left(\frac{\partial B_y}{\partial x}-\frac{\partial B_x}{\partial y}\right)\bigg\rvert_{z=0}, \ee \label{eq:alpha}
where the derivatives are approximated using centered differencing. The uncertainties in $\alpha_0$, which we designate as $\sigma_0$, are derived from the uncertainties in the observed $\bb$ field values. These uncertainties arise from instrumental errors, as well as uncertainties in the inversion process \citep{hoeksemaetal14}, which are propagated to assign uncertainty values for each quantity at every pixel in the SHARP magnetogram \citep{bobraetal14}.

The self-consistency procedure is as follows \citep{wheatlandregnier09}. The CFIT method is first applied to the boundary value problem given by the values of $B_z$ and $\alpha_0$ on the P region (the region where $B_z > 0$) until the NLFFF ``P solution" is obtained. Since the force-free parameter is constant along force-free field lines, the values of $\alpha_0$ from the P region are mapped via this solution to the conjugate magnetic foot points in the N region (the region where $B_z < 0$). These mappings define a new set of values $\alpha_{1}$ in N. CFIT is applied again using the values of $\alpha_0$ on the N region until a second NLFFF result is obtained (the ``N solution"), and the values of $\alpha_0$ from the N region map via this solution to new values $\alpha_{1}$ on the P region. We now have two NLFFF solutions and two complete sets of $\alpha$ values (defined over the entire region, both P and N): $\alpha_{0}$ and $\alpha_{1}$. The uncertainties for the $\alpha_1$ values are obtained by mapping the values of $\sigma_0$ from the P (N) region to the conjugate magnetic foot points in the N (P) region. At each point on the region, we now have two possible sets of values $(\alpha_0, \sigma_0)$ and $(\alpha_{1},\sigma_{1})$.

Bayes' theorem can be used to decide on the most probable value of $\alpha$, given some weighting factor \citep{wheatlandregnier09}. When working with real data, the weighting factor is taken to be the uncertainties $\sigma$ in the values of $\alpha$. The most probable value of $\alpha$ is 

\be \alpha_2=\frac{\alpha_0/\sigma_0^2 + \alpha_1/\sigma_2^2}{1/\sigma_0^2+1/\sigma_1^2},\label{eq:alpha2}\ee
with the corresponding uncertainty

\be \sigma_2=\left(1/\sigma_0^2+1/\sigma_1^2\right)^{-1/2}.\label{eq:sigma2}\ee
This process defines one ``self-consistency cycle" \citep{wheatlandregnier09}. The resulting values of $\alpha_2$ may still be inconsistent with a force-free field, but they are expected to be closer to consistency. The self-consistency cycle is repeated using the force-free parameter values $\alpha_2$ to obtain a new set of values $\alpha_3$, and so on, until the P and N solutions converge to an identical NLFFF solution.

If the boundary values of $\alpha$ on the P and N regions are highly inconsistent with the force-free model, the final self-consistent values of $\alpha$ may be altogether different again. The field lines of the self-consistent field may disagree with structures seen in EUV, and the self-consistent field may have less free energy than that implied by the explosive events produced by the region. These problems were seen in the study of AR 12017 by \cite{kleintetal18}. In such cases, we can consider one set of boundary conditions (\emph{i.e.} P or N) to be favoured over the other, because the corresponding solution reproduces structures of interest. This suggests the idea of weighting the self-consistent solution towards the favoured boundary condition. We can weight the solution by decreasing the uncertainty values of $\alpha$ in the P or N region (that is, $\sigma_0$ or $\sigma_1$ respectively) by some factor. When $\alpha_2$ is calculated using Equation \ref{eq:alpha2}, reducing $\sigma_0$ (or $\sigma_1$) results in $\alpha_2$ being closer to $\alpha_0$ (or $\alpha_1$).

\section{Application to an Analytic Bipole}
    \label{analytic}

In this section we demonstrate the weighting method, applied to an artificially constructed analytic bipole field. 

\begin{figure*}
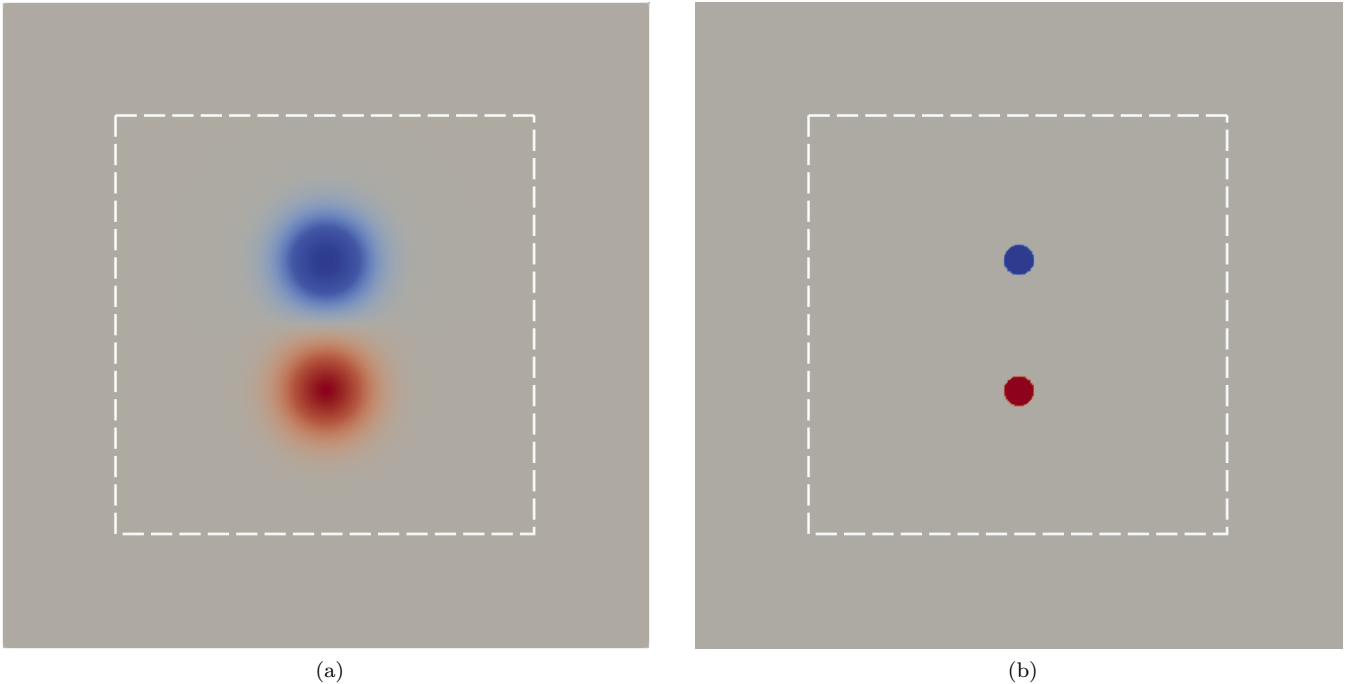

     \subfloat[\label{fig:bz0_b}]{%
      \includegraphics[width=0.48\textwidth]{Bz0_quater.png}
     }
     \hfill
     \subfloat[\label{fig:bz0_a}]{%
      \includegraphics[width=0.48\textwidth]{alpha0_quater.png}
     }
\caption{(a) The boundary conditions for the vertical component of the magnetic field, $B_z(x,y,z=0)$, and (b) the force-free parameter, $\alpha(x,y,z=0)$ of the analytic bipole field, given by Equations \ref{bz0eq} and \ref{eqalphacond}, respectively. Blue denotes positive values, and red denotes negative values. The dashed boxes indicate the cropped field of view of Figure \ref{fig:b} and Figure \ref{fig:alpha}.}
\label{bz0}
\end{figure*}

Figure \ref{bz0} shows the boundary conditions for our calculations. The boundary field $B_z(x,y,z=0)$ is constructed in the region $0\leqslant x \leqslant 1$, $0\leqslant y\leqslant 1$. In dimensionless units, $B_z$ at the bottom boundary of the volume is

\begin{equation}\label{bz0eq}
\begin{split}
B_z(x,y,z=0)&=B_\mathrm{max}\exp[-c_1(x-0.5)^2-c_1(y-0.6)^2]\\
&\quad-B_\mathrm{max}\exp[-c_2(x-0.5)^2-c_2(y-0.4)^2],
\end{split}
\end{equation}
where we set $B_\mathrm{max}$ such that the maximum absolute field strength is 1, and $c_1=c_2=200$. The boundary values of the force-free parameter $\alpha$ (again in dimensionless units) are chosen to be

\begin{equation}\label{eqalphacond}
\alpha(x,y,z=0)= \left\{
\begin{aligned}
&20\textrm{, if }B_z(x,y,z=0)> 0.9,\\
&-20\textrm{, if }B_z(x,y,z=0)< -0.9.
\end{aligned}
\right.
\end{equation}
These boundary values of $\alpha$ are clearly inconsistent with a closed force-free field. A field line traced from the positive polarity to the negative polarity cannot have the same values of $\alpha$ at the two foot points, contradicting Equation \ref{eq:ff2}.

We can apply the self-consistency procedure \citep{wheatlandregnier09} to find a single NLFFF solution. However, to do this we need to assign uncertainties $\sigma$ to the values of $\alpha$. A nominal choice is to choose equal uncertainties, \emph{i.e.} $\sigma=1$ everywhere. We call this the ``equally-weighted" solution. We also consider choosing $\sigma$ to be constant in each polarity, but with a different value. Solutions constructed in this way are weighted towards the polarity with the smaller value of $\sigma$. Note that, strictly speaking, $\sigma$ in this case has no physical meaning, it is simply a weighting factor assigned to $\alpha$.

We use CFIT to calculate the P, N, and self-consistent NLFFF solution in the region $0\leqslant x\leqslant 1$, $0\leqslant y\leqslant 1$, $0\leqslant z\leqslant 0.75$, with the calculation performed on a $400\times 400\times 300$ grid. The P (N) solution is calculated using the $\alpha$ value from the positive (negative) $B_z$ region. The self-consistent solution is calculated using the self-consistency procedure described in Section \ref{method}, for different choices of the uncertainties $\sigma$. 

\begin{figure*}
     \subfloat[\label{fig:b_p}]{%
       \includegraphics[width=0.3\textwidth]{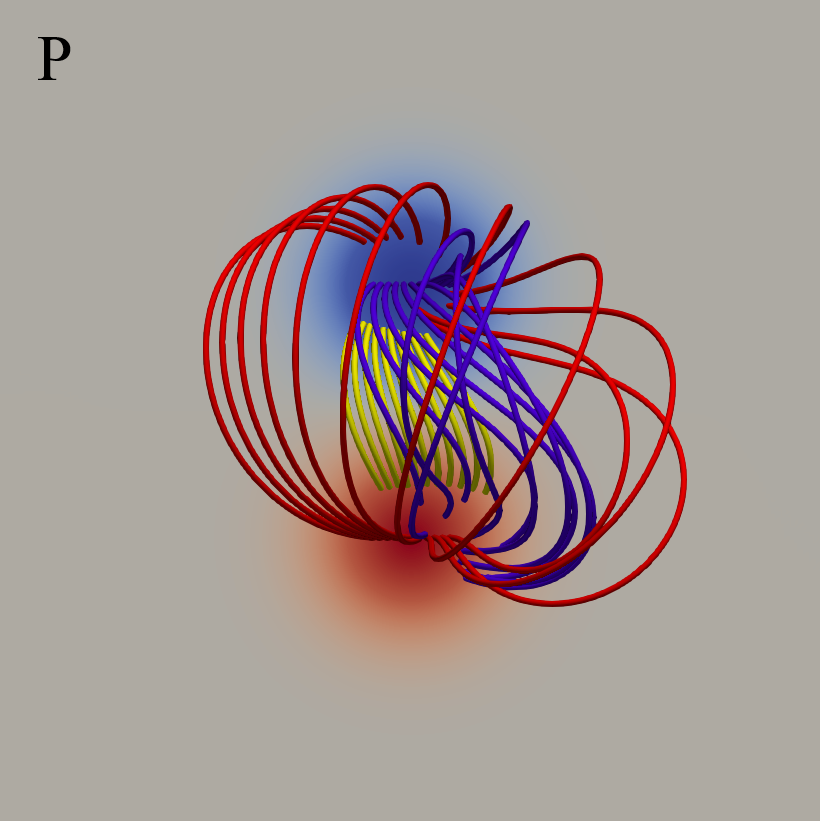}
     }
     \hfill
     \subfloat[\label{fig:b_n}]{%
       \includegraphics[width=0.3\textwidth]{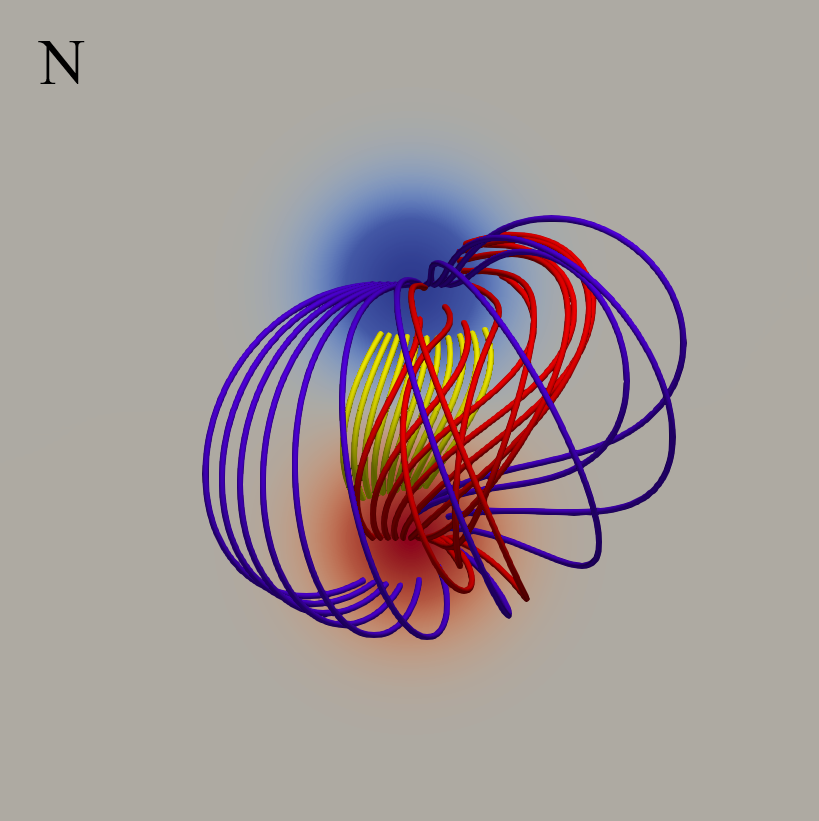}
     }
     
     \subfloat[\label{fig:b_sc}]{%
        \includegraphics[width=0.3\textwidth]{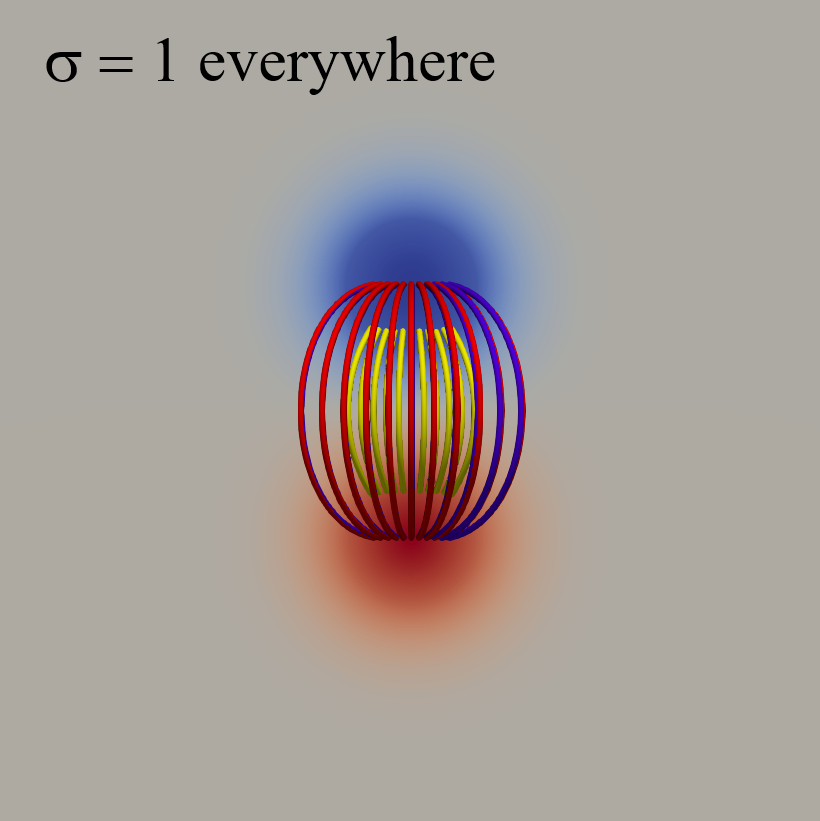}
     }
     \hfill
     \subfloat[\label{fig:b_0p8}]{%
       \includegraphics[width=0.3\textwidth]{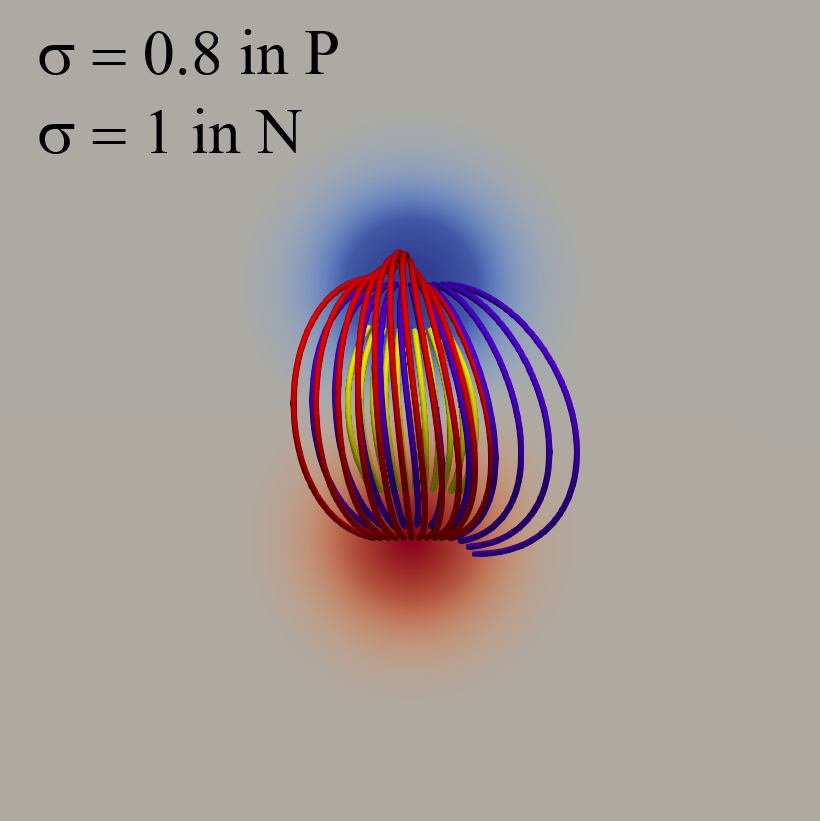}
     }
     
     \centering
     \subfloat[\label{fig:b_0p4}]{%
       \includegraphics[width=0.3\textwidth]{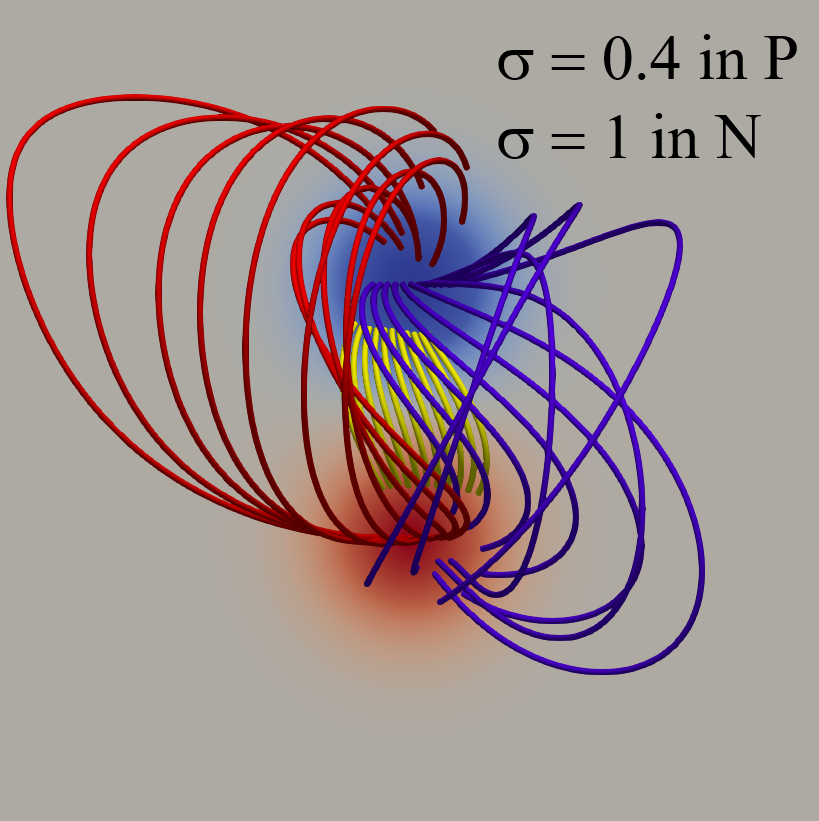}
     }
    
     \caption{Nonlinear force-free field extrapolations of the analytic bipole generated by CFIT, with the boundary conditions given by Equations \ref{bz0eq} and \ref{eqalphacond}. (a) P solution. (b) N solution. (c) Self-consistent solution where the uncertainty value is uniform. (d) Self-consistent solution where the uncertainty value in the P region is 0.8 and the uncertainty in the N region is 1. (e) Self-consistent solution where the uncertainty in the P region is 0.4 and the uncertainty in the N region is 1.}
     \label{fig:b}
   \end{figure*}

Figure \ref{fig:b} shows the magnetic field lines of the solutions calculated by CFIT. Blue (red) field lines are those traced from the positive (negative)  poles, and yellow field lines are selected field lines that pass through $0.45\leqslant x\leqslant 0.55$, $y=0.5$, $z=0.04$. Figure \ref{fig:b_p} shows the P solution, while Figure \ref{fig:b_n} shows the N solution. Because $\alpha$ is positive in the positive pole and negative in the negative pole, as given by Equation \ref{eqalphacond}, the P and N solutions are antisymmetric, \emph{i.e.} the field lines twist in opposite senses. We see in Figure \ref{fig:b_sc} that the equally-weighted ($\sigma=1$ everywhere) self-consistent solution is approximately a potential bipole field, and the self-consistent solution is very different from both the P and N solutions.

Figure \ref{fig:b_0p8} shows the self-consistent CFIT solution obtained by setting $\sigma=0.8$ in the P region, and $\sigma=1$ in the N region. The field lines show some twist in the sense of the P solution (Figure \ref{fig:b_p}). Figure \ref{fig:b_0p4} shows the solution with $\sigma=0.4$ in the P region, and $\sigma=1$ in the N region. The field lines begin to approximate those of the P solution (Figure \ref{fig:b_p}). Figures \ref{fig:b_0p8} and \ref{fig:b_0p4} therefore show how we can make the self-consistent solution more similar to the P solution by reducing $\sigma$ in the P region.

\begin{figure*}
     \subfloat[\label{fig:bipole_alpha_P}]{%
       \includegraphics[width=0.3\textwidth]{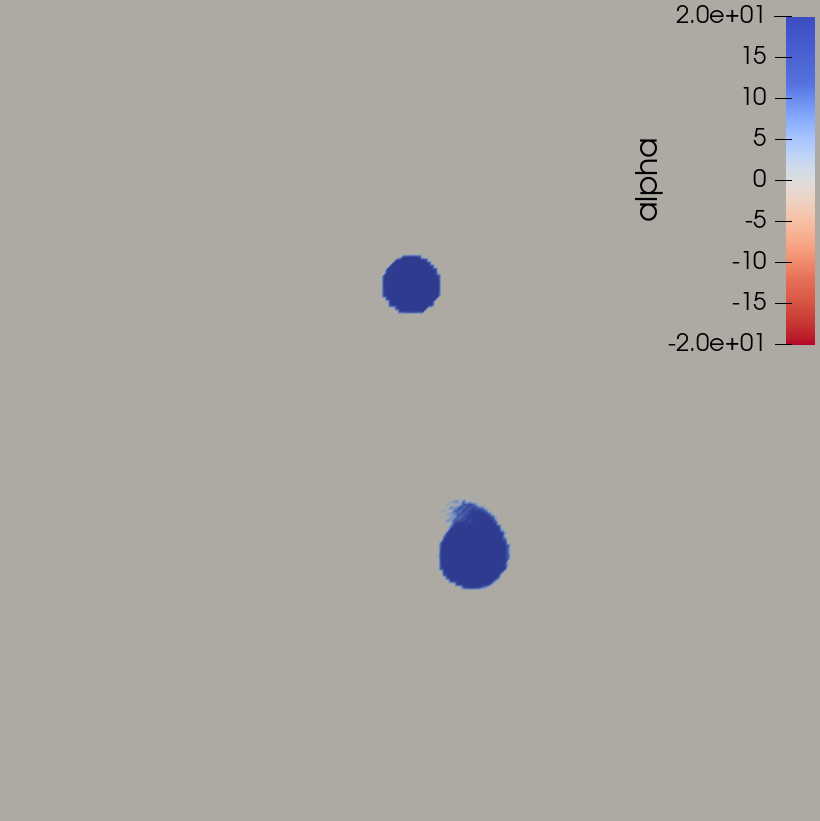}
     }
     \hfill
     \subfloat[\label{fig:bipole_alpha_N}]{%
       \includegraphics[width=0.3\textwidth]{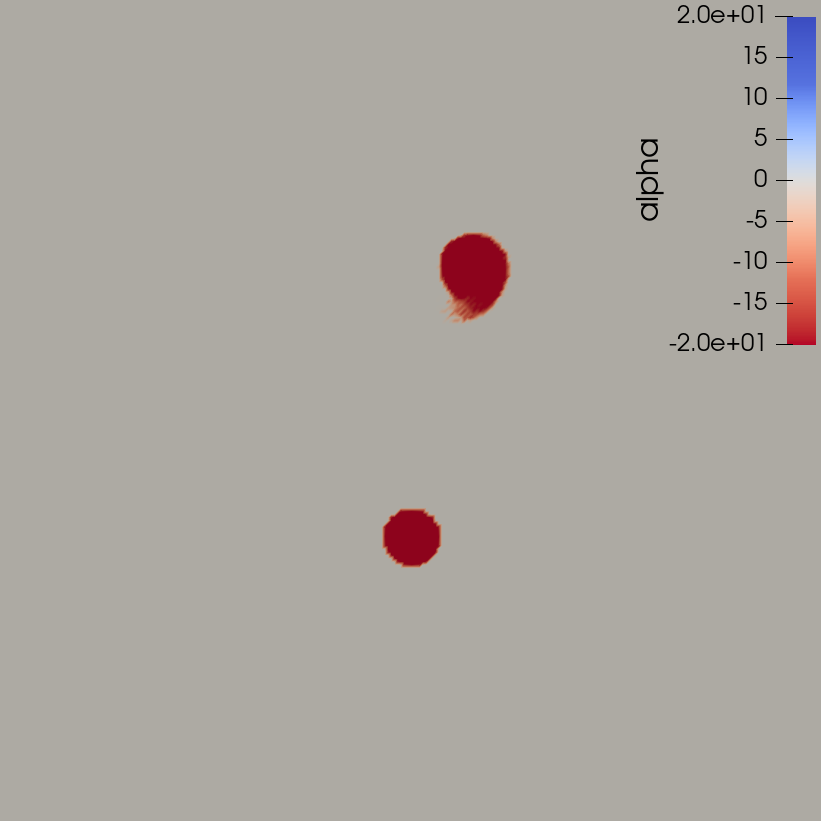}
     }
     
     \subfloat[\label{fig:bipole_alpha_sc}]{%
        \includegraphics[width=0.3\textwidth]{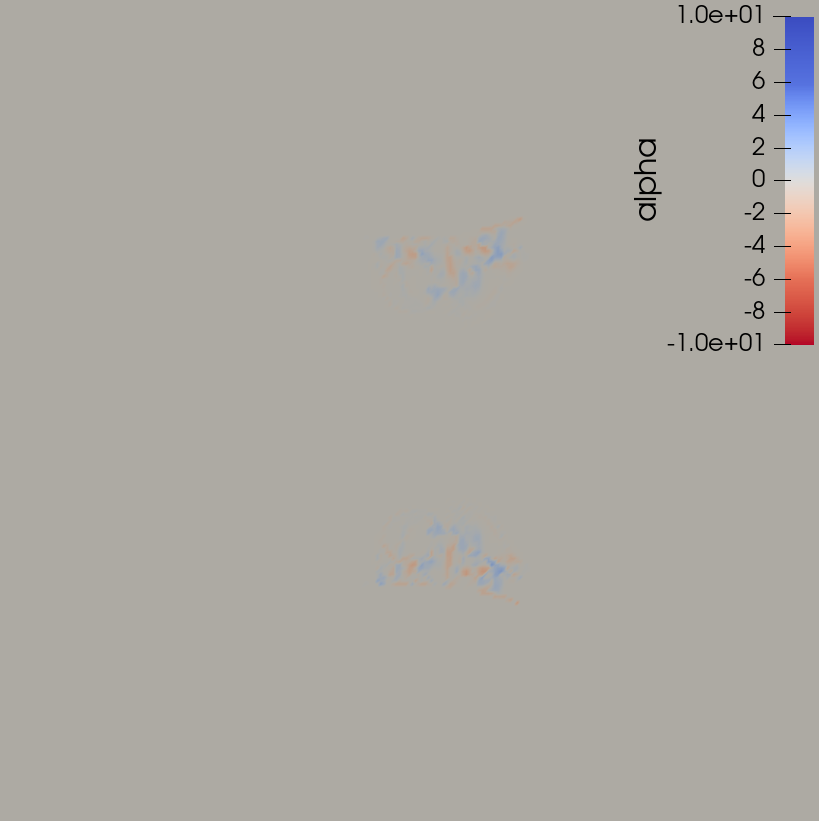}
     }
     \hfill
     \subfloat[\label{fig:bipole_alpha_fP0p8}]{%
       \includegraphics[width=0.3\textwidth]{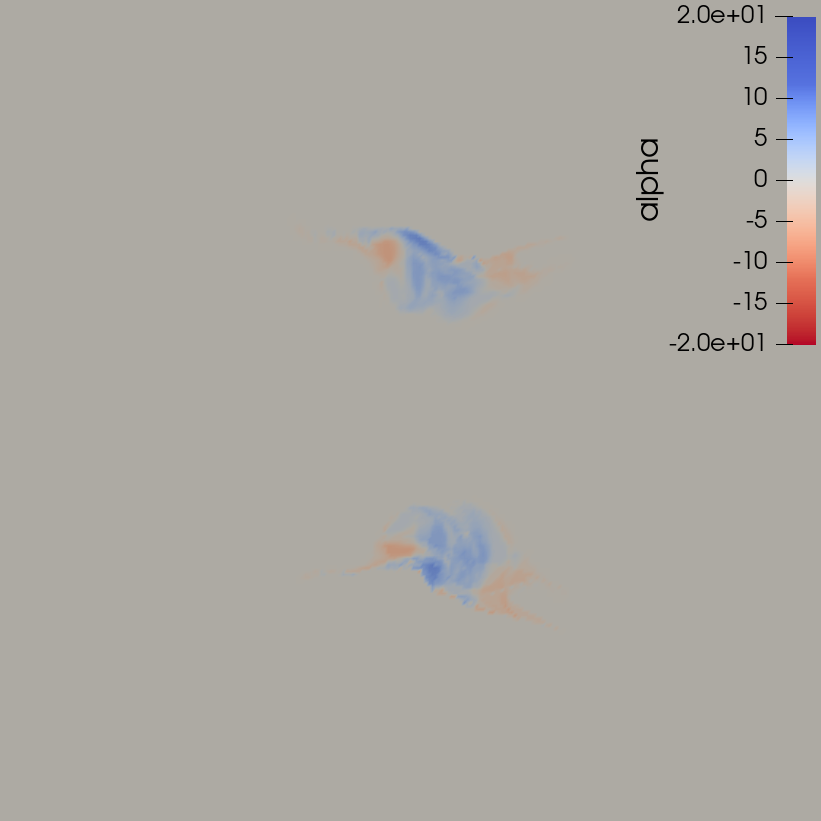}
     }
     
     \centering
     \subfloat[\label{fig:bipole_alpha_fP0p4}]{%
       \includegraphics[width=0.3\textwidth]{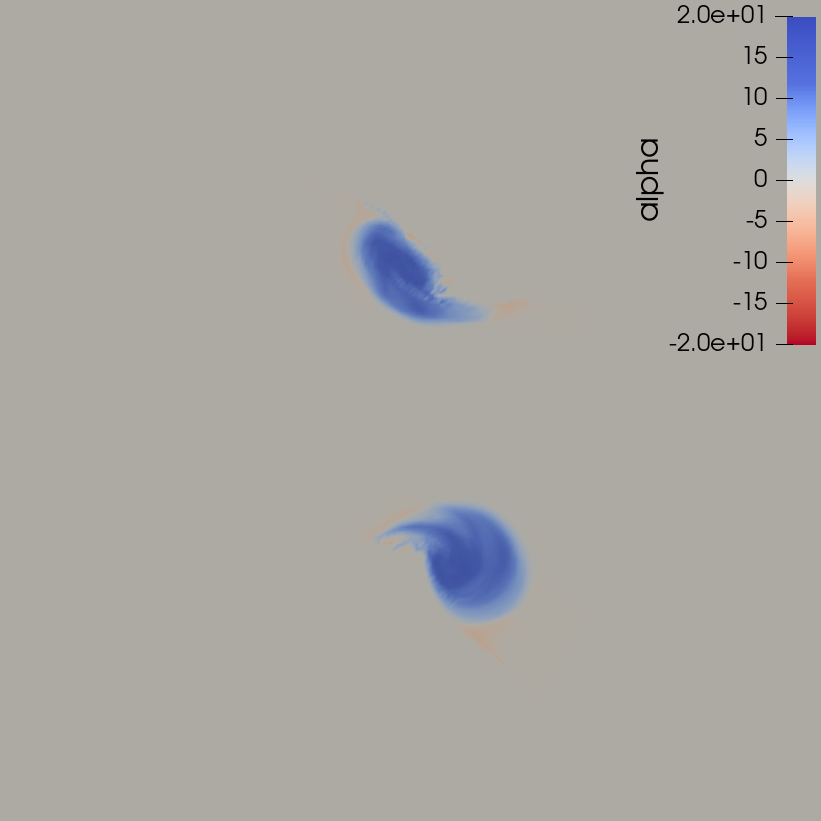}
     }
    
     \caption{The values of $\alpha$ at the lower boundary for the CFIT solutions generated using the boundary conditions given by Equations \ref{bz0eq} and \ref{eqalphacond}. (a) P solution. (b) N solution. (c) Self-consistent solution where the uncertainty value is uniform. (d) Self-consistent solution where the uncertainty value in the P region is 0.8 and the uncertainty in the N region is 1. (e) Self-consistent solution where the uncertainty in the P region is 0.4 and the uncertainty in the N region is 1.}
     \label{fig:alpha}
   \end{figure*}

Figure \ref{fig:alpha} shows the self-consistent values of $\alpha$ at the bottom boundary. Figure \ref{fig:bipole_alpha_P} shows the P solution and Figure \ref{fig:bipole_alpha_N} shows the N solution. In the P (N) solution, field lines traced between the poles are assigned $\alpha=20$ $(-20)$, \emph{i.e.} the value of $\alpha$ in the P (N) region. Figure \ref{fig:bipole_alpha_sc} shows the self-consistent values of $\alpha$ at the bottom boundary for the equally-weighted solution. The self-consistent values of $\alpha$, calculated by successive averaging of the boundary values of the P and N solutions, have absolute values much smaller than 20 (the boundary value of $|\alpha|$ before the self-consistency procedure) everywhere, with positive and negative values of $\alpha$ dispersed in the boundary (note that, for clarity, the colour scale in Figure \ref{fig:bipole_alpha_sc} is different from the other panels of Figure \ref{fig:alpha}). Figure \ref{fig:bipole_alpha_fP0p8} shows the values of $\alpha$ with $\sigma=0.8$ in the P region, and $\sigma=1$ in the N region. The values and distribution of $\alpha$ are now closer to the P solution, although the values are much smaller. Figure \ref{fig:bipole_alpha_fP0p4} shows the values of $\alpha$ with $\sigma=0.4$ in the P region, and $\sigma=1$ in the N region. The values and distribution of $\alpha$ approximate the P solution, with some small negative values of $\alpha$ remaining.



\section{Application to AR 12017}
    \label{ar12017}
      
In this section, we apply our weighting method to the NLFFF reconstruction of the coronal field of active region AR 12017 for 2014 March 29, 17:36 UT.

On 2014 March 29, 17:48 UT, AR 12017 produced an X1 flare which was simultaneously observed and recorded by a number of instruments and observatories, such as the \emph{Interferometric BIdimensional Spectropolarimeter} (IBIS) instrument at the Dunn Solar Telescope, the \emph{Hinode} spacecraft, and the RHESSI spacecraft \citep{kleintetal15}.

\cite{yangguoding16} generated NLFFF reconstructions of AR 12017 using the optimization method \citep{wheatlandsturrockroumeliotis00,wiegelmann04}, and \cite{woodsetal18} reconstructed the field using the magnetohydrodynamic relaxation method of \cite{inoueetal14} and \cite{inoue16}. \cite{kleintetal18} presented a reconstruction of the field using CFIT, with the self-consistency procedure. The results of the optimization and relaxation methods, as well as AIA EUV observations, clearly show a filament present near the neutral line which erupts during the flare. However, the self-consistent CFIT NLFFF solution does not contain a flux rope structure at the neutral line \citep{kleintetal18}. The P solution does show a highly-twisted flux rope, which is absent in the N solution, and correspondingly, is absent in the self-consistent solution. Hence we consider applying the weighted self-consistent reconstruction procedure, with the boundary conditions on $\alpha$ weighted towards the positive polarity.

The relevant vector magnetogram boundary data for our NLFFF extrapolation are the SHARP data for AR 12017, at 17:36 UT on 2014 March 29. The SHARP patch is $642\times 504$ pixels, where each pixel is $0.502\times 0.502$ arcsec. We approximate the SHARP magnetic field data, given in Lambert equal-area projection, as field values on a Cartesian grid as explained in Section \ref{method}, and solve for the magnetic field in a $642\times 504\times 200$ Cartesian box. The vertical size is chosen to capture as much large-scale structure as possible without unduly slowing down the calculation. For convenience, we work in dimensionless quantities. We use the width of the solution box as the length scale, such that the solution box covers the region $0\leqslant x\leqslant 1$, $0\leqslant y\leqslant 0.785$, and $0\leqslant z \leqslant 0.31$. We normalize the vector magnetic field strength values to the maximum of the absolute value of $B_z$ at $z=0$ (in this case, $|B_z|_\mathrm{max} = 2.649\times 10^3$ G). The boundary values of $\alpha$ are calculated using Equation \ref{eq:alpha}, and the associated uncertainties are propagated from the uncertainties of $\bb$ given in the SHARP data \citep{bobraetal14,hoeksemaetal14}, to give a map of the uncertainties $\sigma$ in $\alpha$ that we use in the self-consistency cycles. The values of $\alpha$ and $\sigma$ are non-dimensionalized by multiplying by the length scale. To reduce the influence of unphysical currents, we set the current to zero in pixels where the signal-to-noise ratio of the current is less than one or where $|B_z|<0.05$.

\begin{figure*}
     \subfloat[\label{fig:ar_p}]{%
       \includegraphics[width=0.3\textwidth]{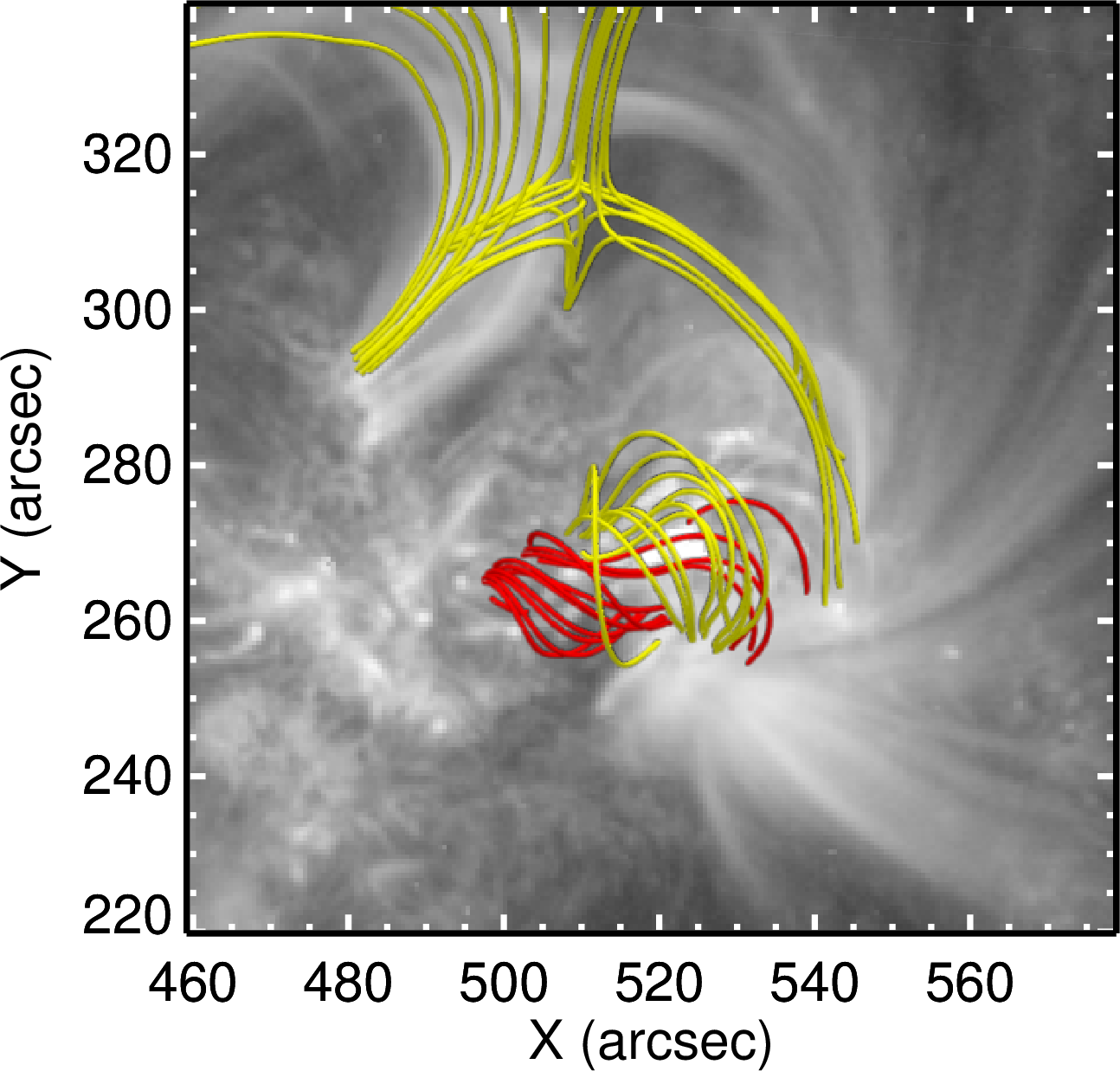}
     }
     \hfill
     \subfloat[\label{fig:ar_n}]{%
       \includegraphics[width=0.3\textwidth]{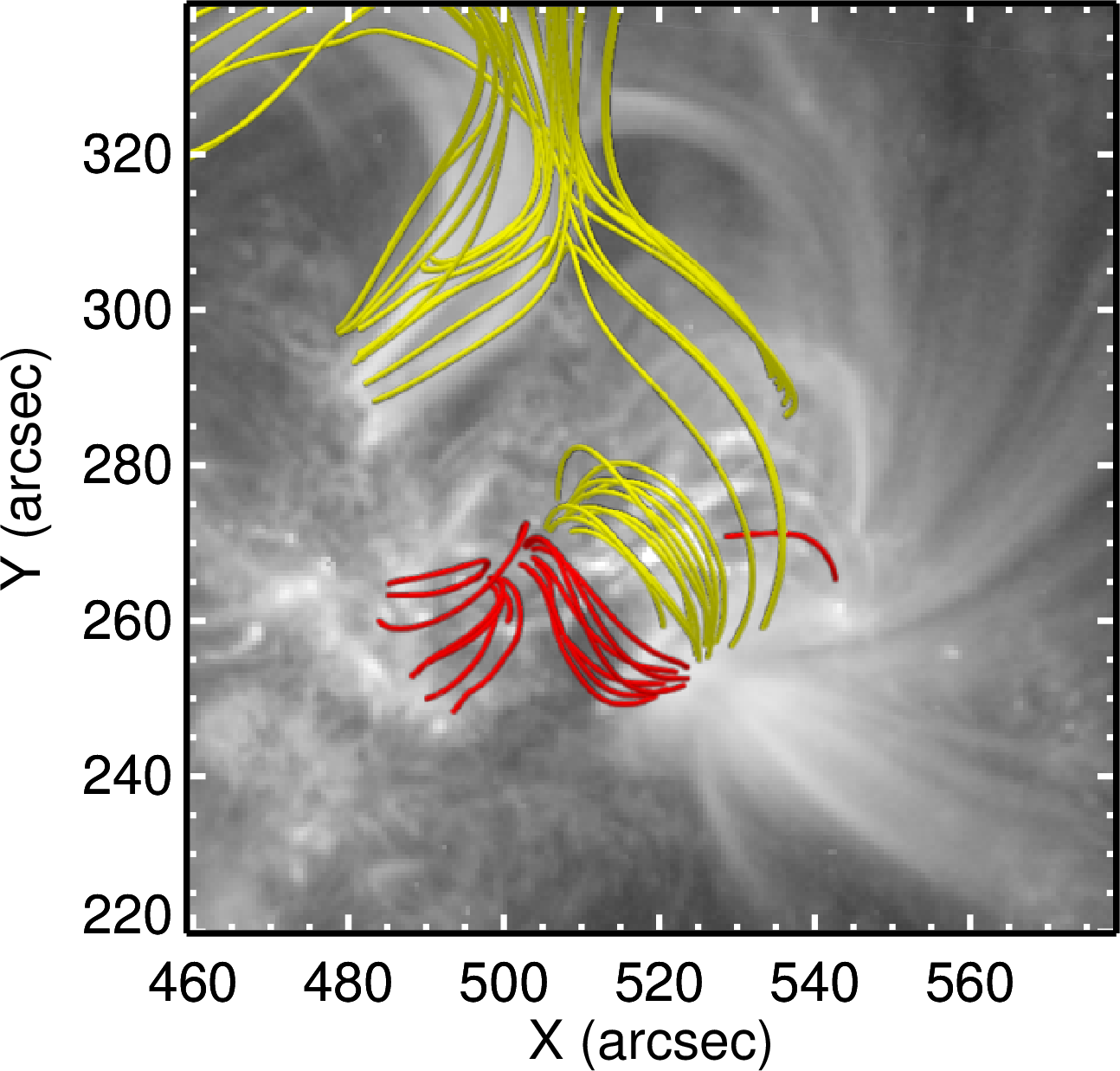}
     }
    
     \centering
     \subfloat[\label{fig:ar_aia}]{%
       \includegraphics[width=0.3\textwidth]{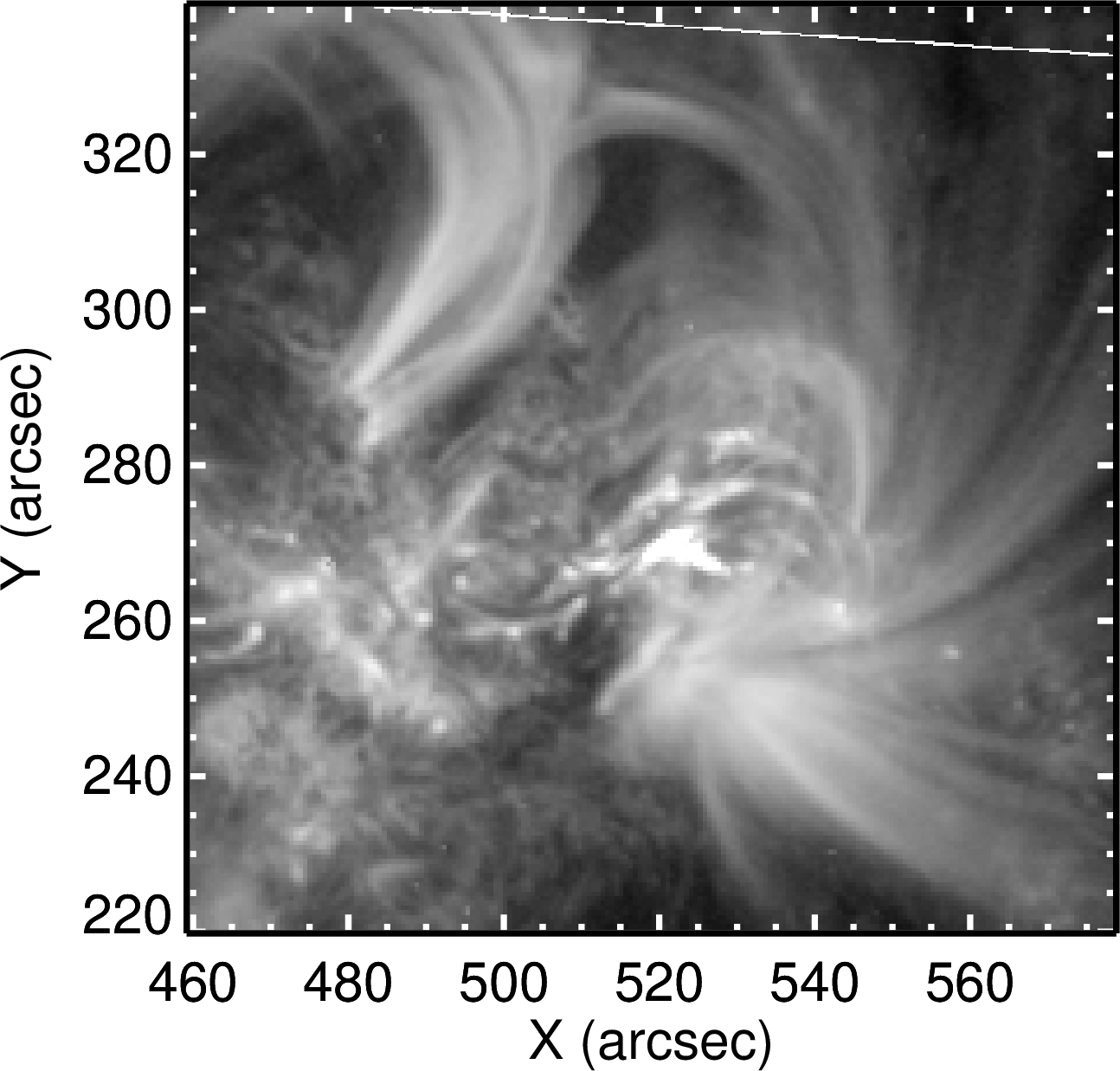}
     }

     \subfloat[\label{fig:ar_sc}]{%
       \includegraphics[width=0.3\textwidth]{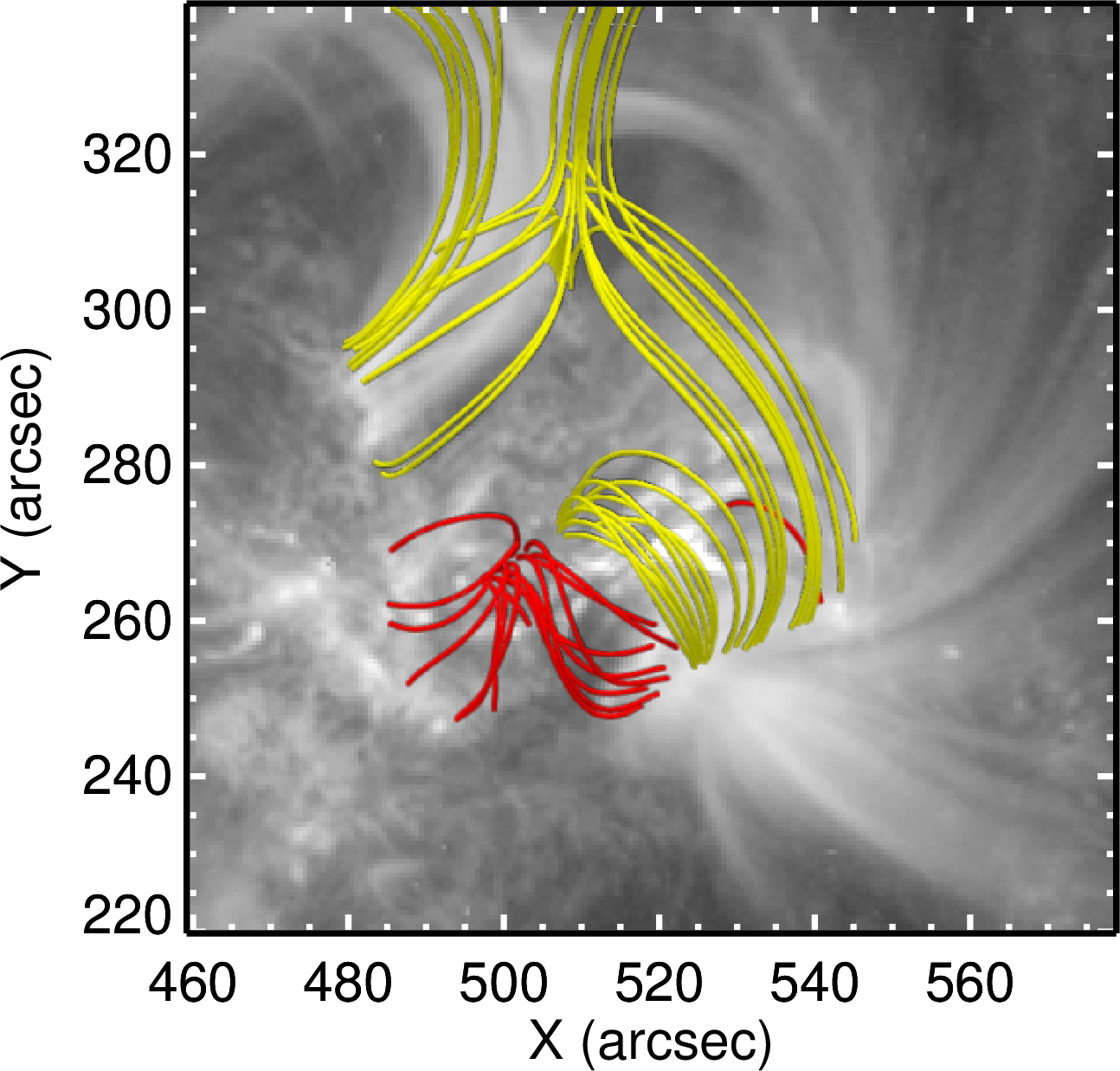}
     }
     \hfill 
     \subfloat[\label{fig:ar_fP0p1}]{%
       \includegraphics[width=0.3\textwidth]{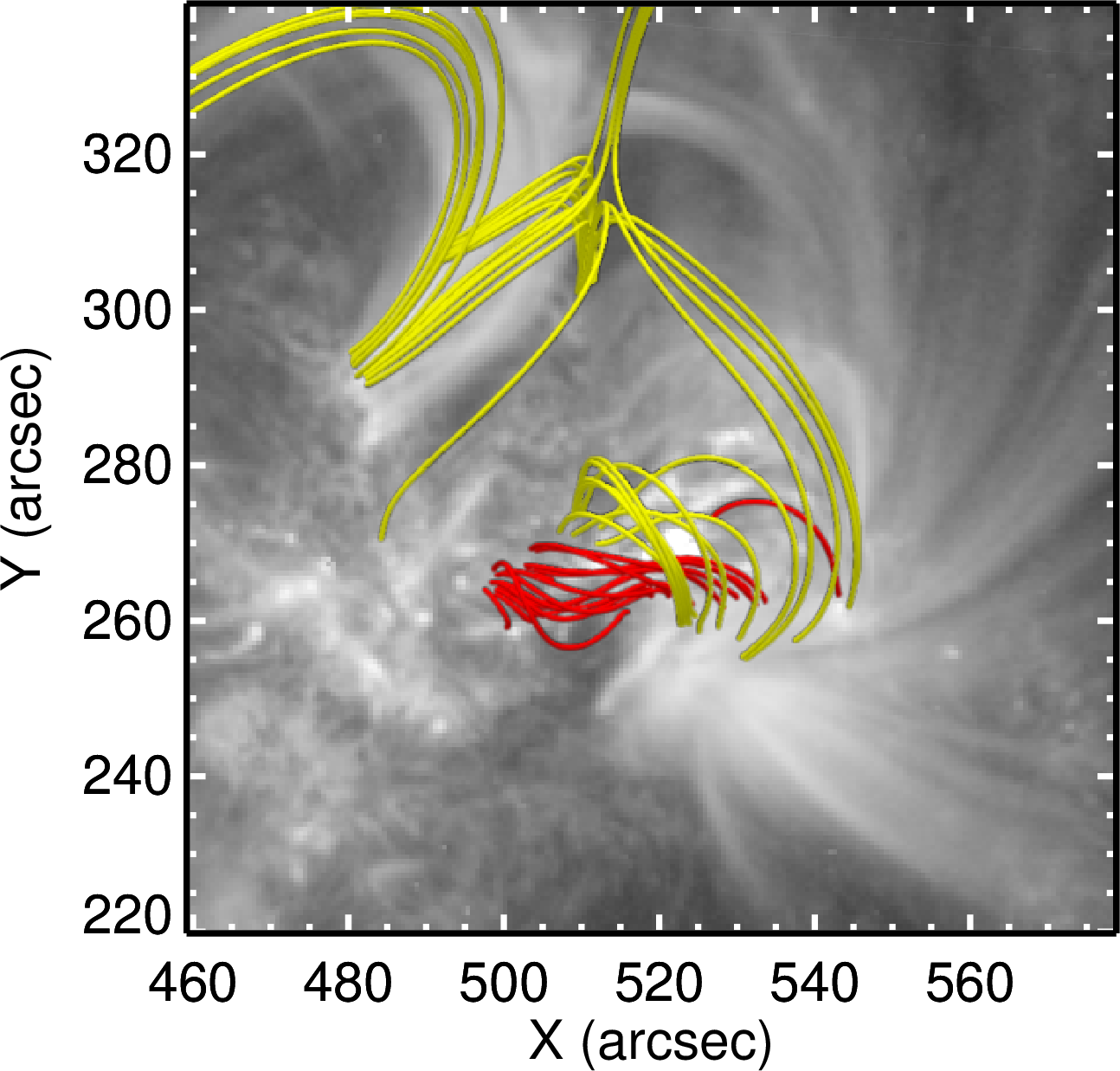}
     }
     

     \caption{The nonlinear force-free magnetic field solutions obtained from the CFIT code and overlaid onto the contemporaneous 171 {\AA} AIA image. The field extrapolations are derived from the 2014 March 29, 17:36 UT, SHARP vector magnetogram data for AR 12017. The panels show (a) the P solution, (c) the N solution, (c) the 171 {\AA} AIA image, (d) the (unweighted) self-consistent solution, and (e) the self-consistent solution from a boundary condition weighted towards the P solution.}
     \label{fig:ar}
   \end{figure*}

Figure \ref{fig:ar} shows the results of the CFIT NLFFF extrapolations, along with the 171 {\AA} EUV AIA image of the same region for comparison (Figure \ref{fig:ar_aia}). The P solution, Figure \ref{fig:ar_p}, shows a highly-twisted flux rope, associated with high values of $\alpha$ near the neutral line, which coincides with the filament structure seen in the AIA image. On the other hand, the N solution, Figure \ref{fig:ar_n}, does not show such a structure. The self-consistent solution, Figure \ref{fig:ar_sc}, has lost the high values of $\alpha$ present in the P boundary condition, and appears as a sheared rather than a twisted field. These results are consistent with \cite{kleintetal18}.

Because we want to generate a self-consistent solution which preserves the high values of $\alpha$ in the P solution (in particular associated with the twisted flux rope), we weight the boundary condition towards the P solution by multiplying $\sigma$ in the P region at locations where $50<|\alpha|<150$ (in dimensionless units) by a factor of 0.1. We arrive at these upper and lower limits on $|\alpha|$ by visually comparing the P solution to the AIA images. We inspect the foot points of the flux rope in the P solution and we conclude that they are associated with $|\alpha|\gtrsim 50$ (or $|\alpha| \gtrsim 2.14\times 10^{-7}$ m$^{-1}$ in dimensional units). In addition, because we want to limit the effects of high, unphysical values of $\alpha$ (which can prevent CFIT from converging to a self-consistent solution), we set an upper limit of $|\alpha|<150$ (or $|\alpha| < 6.42\times 10^{-7}$ m$^{-1}$). More systematic ways of identifying $\alpha$ values are available, such as that proposed by \cite{malanushenkolongcopemckenzie09}. For our purposes, visual inspection and comparison suffice. We present the result in Figure \ref{fig:ar_fP0p1}. The new self-consistent solution clearly preserves the flux rope in the P solution (Figure \ref{fig:ar_p}), which is absent in the unweighted self-consistent solution (Figure \ref{fig:ar_sc}). 

We can calculate the free energy of our solutions, defined as the difference between total energy $E$ of the extrapolated NLFFF and the energy $E_0$ of the potential component of the NLFFF. The free energy of the P solution is $8.1\times 10^{31}$ erg (corresponding to $E/E_0 = 1.18$), the free energy of the N solution is $2.0\times 10^{31}$ erg ($E/E_0=1.05$), the free energy of the unweighted self-consistent solution is $5.4\times 10^{30}$ erg ($E/E_0=1.015$), and the free energy of the weighted self-consistent solution is $3.4\times 10^{31}$ erg ($E/E_0=1.09$). The unweighted self-consistent solution has the least free energy of these four solutions, even less than the free energy of the N solution. The weighted solution has more free energy than the N solution, but less free energy than the P solution.

We can also characterize the fields using relative helicity $H_\mathrm{r}$ \citep{bergerfield84,finnantonsen85}. Using the definition given by, \emph{e.g.}, \cite{demoulinberger03}, we separate the magnetic field $\bb$ in the computational volume $V$ into the potential field $\vec{B}_\mathrm{p}$ and the current-carrying field $\vec{B}_\mathrm{c}$

\be \bb=\vec{B}_\mathrm{p}+\vec{B}_\mathrm{c},\ee
where

\be \bb\cdot\en|_{\partial V} = \vec{B}_\mathrm{p} \cdot\en|_{\partial V},\ee
where $\partial V$ is the boundary surface and $\en$ is the unit vector normal to $\partial V$, and we write

\be H_\mathrm{r} = \int_V (\avec + \avec_\mathrm{p})\cdot (\bb-\vec{B}_\mathrm{p}) dV.\ee
For the NLFFF presented here, $H_\mathrm{r}=1.91\times 10^{44}$ Mx$^2$ for the unweighted self-consistent solution and $H_\mathrm{r}=9.91\times 10^{44}$ Mx$^{2}$ for the weighted solution. The weighted solution has larger currents than the unweighted solution and consequently higher relative helicity.

\begin{figure*}
     \subfloat[\label{fig:ar_bz_f}]{%
       \includegraphics[width=0.3\textwidth]{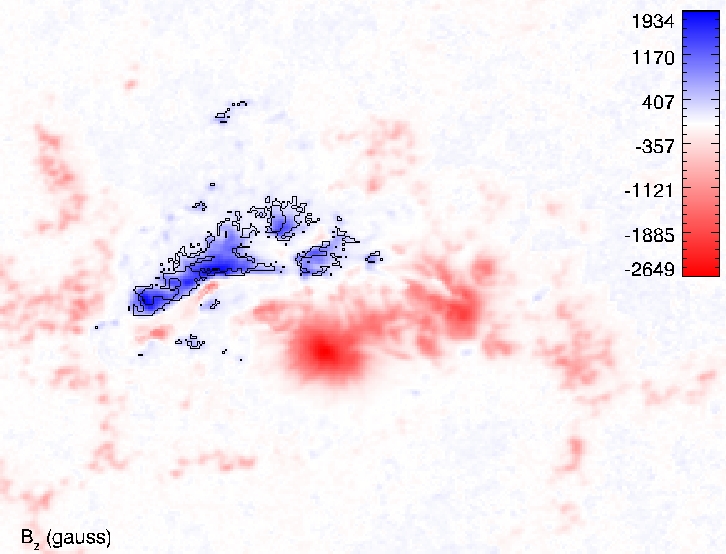}
     }
     \hfill
    \subfloat[\label{fig:ar_sig0v}]{%
       \includegraphics[width=0.3\textwidth]{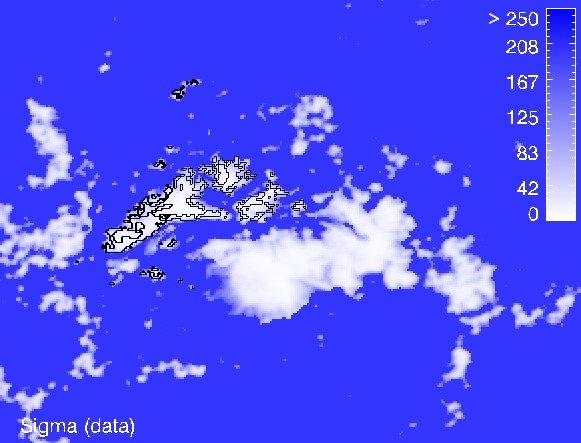}
     } 
     
     \subfloat[\label{fig:ar_jz0}]{%
       \includegraphics[width=0.3\textwidth]{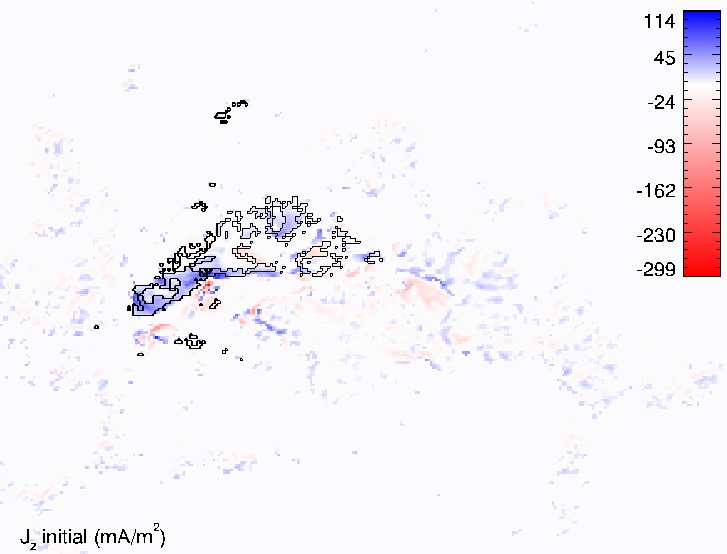}
     }
    \hfill
     \subfloat[\label{fig:ar_jzf}]{%
       \includegraphics[width=0.3\textwidth]{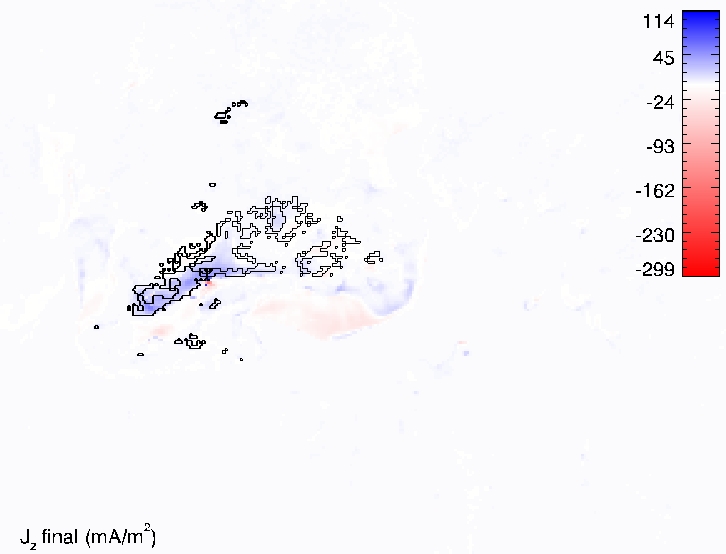}
     }

    \centering
     \subfloat[\label{fig:ar_bz_lines}]{%
       \includegraphics[width=0.3\textwidth]{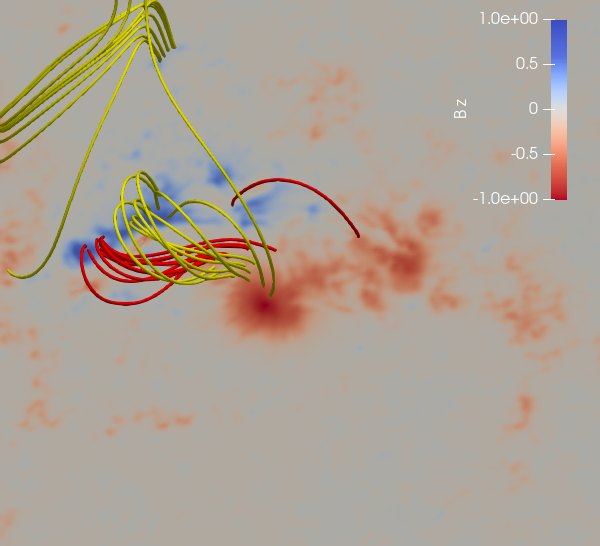}
     }

     \caption{Maps of the vertical magnetic field values ($B_z$) and vertical current density ($J_z$) on the plane tangent to the solar surface. Panel (a) shows $B_z$, (b) shows the absolute value of the uncertainty $\sigma$ in $\alpha$ from the data, (c) shows the initial boundary conditions on $J_z$ (before the self-consistency procedure), (d) shows $J_z$ corresponding to the weighted self-consistent NLFFF solution, and (e) shows selected field lines of the weighted self-consistent NLFFF solution, superimposed on the $B_z$ map. In panels (a)--(d), the contours enclose the regions where we reduce $\sigma$.}
     \label{fig:ar_z}
   \end{figure*}

Figure \ref{fig:ar_z} shows maps of $B_z(x,y,z=0)$ in units of gauss, the uncertainty $\sigma$ in dimensionless units, the current density $J_z(x,y,z=0)$ in units of mA m$^{-2}$, and a selection of extrapolated NLFFF field lines. Figure \ref{fig:ar_bz_f} shows the boundary condition on $B_z$, with the weighted areas outlined. Figure \ref{fig:ar_sig0v} shows a map of the absolute value of the dimensionless uncertainty $\sigma$ in $\alpha$, with the weighted areas outlined, saturated to 250 for clarity. Figure \ref{fig:ar_jz0} and Figure \ref{fig:ar_jzf} show the values of the boundary data and the self-consistent vertical current densities $J_z$ respectively, with the weighted areas outlined.  Weak-field areas, where $|B_z|<0.05 |B_z|_\mathrm{max}$, are assigned $\alpha=0$ and a maximum value of $\sigma$. We see that, while the distribution of the current has been smoothed out, the weighting procedure preserves areas of strong current density. We show a plot of selected field lines superimposed on the $B_z$ map in Figure \ref{fig:ar_bz_lines}, for comparison. Figure \ref{fig:ar_bz_lines}, in conjunction with Figure \ref{fig:ar_jzf}, show that the flux rope is associated with the high-$\alpha$ P region where we reduce $\sigma$. Note that the point of view in Figure \ref{fig:ar_z} looks straight down the plane tangent to the photosphere, not tilted to the image plane like in Figure \ref{fig:ar_fP0p1}.

%
%
%
%
%

\section{Conclusions} 
      \label{conclusions}

In this article, we present a modification to the self-consistency procedure implemented using the Grad-Rubin code CFIT \citep{wheatlandregnier09}. It is desirable to arrive at a self-consistent NLFFF solution for coronal magnetic fields using the boundary data given by vector magnetogram observations. In general, the values of $\alpha$ in the P and N magnetic polarities are not consistent with a force-free field. The self-consistency procedure of \cite{wheatlandregnier09} solves this problem by constructing P and N solutions and taking uncertainty-weighted averages of the boundary values of $\alpha$ from the solutions, iteratively, to arrive at a self-consistent solution. We demonstrate this with an analytic bipole, which has inconsistent boundary values of $\alpha$ on the P and N poles by construction. The unweighted self-consistent solution using these boundary conditions is different from both the P and N solutions, and is close to a potential field. We show that by reducing the uncertainties $\sigma$ in the values of $\alpha$ in the P region, we can skew the self-consistent solution towards the P solution.


We apply this procedure to active region AR 12017. The P and N solutions obtained by CFIT from the 2014 March 29, 17:36 UT time step are inconsistent with each other. When the P and N boundary conditions differ significantly, the resulting self-consistent NLFFF solution may, like the simple bipole case, be close to a potential field, and not provide an accurate model for the true coronal field, based \emph{e.g.} on comparison of field lines with structures observed in EUV images. The P and N solutions for this time step demonstrate this. Because the P solution has field lines which are more consistent with the observations, and contains greater free energy, we weight the boundary conditions towards the P region. To achieve this, we identify the foot points of the flux ropes in the P solution and multiply the values of $\sigma$ at the foot points by a factor of 0.1. We calculate the new self-consistent solution and, like the P solution, it shows the flux rope evident in EUV.


In this article, we show how a simple modification to the Grad-Rubin CFIT code can achieve self-consistent NLFFF solutions using boundary conditions derived from readily available HMI photospheric vector magnetograms, taking into account the errors in the tangential field strength measurements. We show how visible field structures can be used as an additional constraint in a Grad-Rubin NLFFF extrapolation. In the example presented in this article, the choice of which pixels to be weighted is based on visual inspection. In future work, we will investigate an automated approach, similar to those proposed by \cite{malanushenkolongcopemckenzie09} and \cite{malanushenkoetal14}. 



\phantom{+}

This work was funded in part by an Australian Research Council Discovery Project (DP160102932). The authors thank Don Melrose for helpful comments.

The authors declare that they have no conflicts of interest.


\bibliographystyle{spr-mp-sola}
\bibliography{solar} 

\end{document}